\documentclass[
preprint,
superscriptaddress,
 amsmath,amssymb,
 aps,
prb,
]{revtex4-2}

\usepackage{txfonts}
\usepackage{color}
\usepackage{ulem} 
\usepackage{graphicx}
\usepackage{dcolumn}
\usepackage{bm}
\usepackage{braket}

\begin{document}

\preprint{preprint}

\title{Single-site quadrupolar Kondo effect in a diluted non-Kramers doublet system\\ Y$_{1-x}$Pr$_x$Ir$_2$Zn$_{20}$ for $x = 0.028$ viewed from magnetization}

\author{Yu Yamane}
  \email{yamane.yu.es@osaka-u.ac.jp}
\altaffiliation[Present Address:]{Graduate School of Engineering Science, Osaka University, Toyonaka, Osaka 560-8531, Japan}
 \affiliation{Department of Quantum Matter, Graduate School of Advanced Science and Engineering, Hiroshima University, Higashi-Hiroshima 739--8530, Japan}
 \affiliation{Department of Material Science, Graduate School of Science, University of Hyogo, Kamigori, Hyogo 678-1297, Japan}

\author{Takahiro Onimaru}%
 \affiliation{Department of Quantum Matter, Graduate School of Advanced Science and Engineering, Hiroshima University, Higashi-Hiroshima 739--8530, Japan}

\author{Yasuyuki Shimura}%
 \affiliation{Department of Quantum Matter, Graduate School of Advanced Science and Engineering, Hiroshima University, Higashi-Hiroshima 739--8530, Japan}

\author{Suguru Tsuda}%
 \affiliation{Department of Quantum Matter, Graduate School of Advanced Science and Engineering, Hiroshima University, Higashi-Hiroshima 739--8530, Japan}

\author{Kazunori Umeo}%
 \affiliation{Department of Low Temperature Experiment, Integrated Experimental Support / Research Division, N-BARD,
Hiroshima University, Higashi-Hiroshima 739--8526, Japan}

\author{Toshiro Takabatake}%
 \affiliation{Department of Quantum Matter, Graduate School of Advanced Science and Engineering, Hiroshima University, Higashi-Hiroshima 739--8530, Japan}

\date{\today}

%
%


\begin{abstract}
A diluted non-Kramers doublet system Y$_{1-x}$Pr$_x$Ir$_2$Zn$_{20}$ is a promising candidate for exhibiting single-site quadrupolar (two-channel) Kondo effect. We have measured temperature-dependent magnetization of a sample for $x$ = 0.028 down to 0.1 K at various constant magnetic fields to extract the characteristic behaviors due to the quadrupolar Kondo effect.
The Curie--Weiss fit to the magnetic susceptibility between 50 and 300 K yields a negative paramagnetic Curie temperature of $-$7.0 K, indicative of  on-site antiferromagnetic interaction.
The magnetization divided by magnetic field, $M(T)/B$, at $B = 0.5$ T is saturated to a constant value below 3 K.
On the contrary, in higher magnetic fields of $B$ = 1 and 2 T, $M(T)/B$ exhibits $-$ln$T$ dependence from 1 to 0.1 K, which temperature dependence is consistent with that of the quadrupolar susceptibility detected from ultrasonic measurements.
The $-$ln$T$ form of $M(T)/B$ appears to reflect the behavior of quadrupolar susceptibility, through the magnetic-field-induced magnetic moment due to the mixing of the non-Kramers doublet and crystalline-electric-field excited states of Pr$^{3+}$.
In $B$ = 4 T, $M(T)/B$ approaches a constant on cooling below 0.3 K, because of the quenching of the quadrupoles in the field-induced singlet ground state.
\end{abstract}


\maketitle

\section{Introduction}

Since the concept of multi-channel Kondo effect was first proposed in 1980 by Nozi\'{e}res and Blandin \cite{Nozieres_1980}, it has been a long standing task for experimentalists to find a system in which this concept is realized.
One possible candidate is a non-Fermi liquid (NFL) state arising from the (single-site) two-channel quadrupolar Kondo effect, which is ascribed to overscreening of a localized electric quadrupole by two equivalent conduction bands \cite{Cox_1987}.
Thereby, the NFL behaviors are expected to manifest themselves at low temperatures; magnetic specific heat divided by temperature $C_{\mathrm{m}}/T \propto -$ln$T$, quadrupolar susceptibility $\chi_{\mathrm{Q}} \propto -$ln$T$, electrical resistivity $\rho \propto 1 \pm  A\sqrt{T}$ ($A$ is a constant), and a half of entropy of $R$ln2 ($R$ is the gas constant) remains at $T = 0$ \cite{Tsvelick_1985, Sacramento_1991, Affleck_1995, Cox_1998}.
So far, much experimental effort has been conducted in diluted systems of $5f^2$ (U$^{4+}$) and $4f^2$ (Pr$^{3+}$) systems such as Th(U)Be$_{13}$, Y(U)Pd$_3$, Th(U)Ru$_2$Si$_2$, and La(Pr)Pb$_3$ \cite{Aliev_1994, Seaman_1991, Amitsuka_1994, Kawae_2006}, whose crystalline electric field (CEF) ground states are possibly non-Kramers doublets carrying active quadrupoles.
The quadrupolar Kondo effect, however, has not been experimentally established yet.
For the dilute U alloys, the itinerant character of the $5f^2$ electrons in the U-ion makes the CEF levels blurred.
For La(Pr)Pb$_3$, a small amount of filamentary impurity Pb conceals the intrinsic temperature variation of $\rho$.

In this work, we have focused on a cubic diluted Pr compound Y$_{1-x}$Pr$_x$Ir$_2$Zn$_{20}$.
In the whole range of $0 < x \leq 1$, the CEF ground state of the Pr$^{3+}$ ion located in a cubic $T_d$ symmetry is the non-Kramers $\Gamma_3$ doublet possessing the electric quadrupoles \cite{Onimaru_2011, Ishii_2011, Iwasa_2013, Yamane_2018a, Yanagisawa_2019}.
The CEF scheme for $x = 1$ has been determined by inelastic neutron scattering experiments to be  $\Gamma_3$ doublet --  $\Gamma_4$ triplet (28 K) --  $\Gamma_1$ singlet (66 K) --  $\Gamma_5$ triplet (68 K) \cite{Iwasa_2013}.
The $x=1$ compound shows an antiferroquadrupolar (AFQ) order at $T_{\mathrm{Q}} = 0.11$ K, below which a superconducting transition occurs at $T_{\mathrm{c}} = 0.05$ K \cite{Onimaru_2010, Onimaru_2011}.
On cooling from 1 K to $T_{\mathrm{Q}} = 0.11$ K, $C_{\mathrm{m}}(T)/T$ exhibits the $-$ln$T$ dependence and $\rho(T)$ shows a downward curvature, suggesting formation of a quadrupolar Kondo lattice \cite{Onimaru_2016, Tsuruta_1999, Tsuruta_2015}.
In the dilute range of $x \leq 0.44$, $C_{\mathrm{m}}(T)/T$ and the differential electrical resistivity, $\Delta\rho(T) = \rho(T) - \rho$(3 K), show the NFL behaviors of $C_{\mathrm{m}}(T)/T \propto -$ln$T$ and $\rho(T) \propto 1 + A\sqrt{T}$ below a characteristic temperature $T_0$ defined as the temperature where the magnetic entropy reaches $0.75R$ln2 \cite{Yamane_2018b}.
If $T_0$ is regarded as the Kondo temperature $T_{\mathrm{K}}$ for the two-channel Kondo model \cite{Sacramento_1991}, the NFL behaviors in  $C_{\mathrm{m}}(T)/T$ and  $\Delta\rho(T)$ can be scaled by $T_{\mathrm{K}}$ \cite{Yamane_2018b}.
This scaling suggests that the NFL behaviors originate from the single-site effect of Pr$^{3+}$.
In addition, the susceptibility of the overscreened quadrupoles have been detected by ultrasonic measurements \cite{Yanagisawa_2019}.
In the magnetic field range of $0 \leq B \leq 2$ T, the elastic constant $(C_{11}-C_{12})/2$, which is identical to the $\Gamma_3$-type quadrupolar susceptibility, exhibits the ln$T$ dependent softening below $T_0$.
For $B > 2$ T, $(C_{11}-C_{12})/2$ reaches a constant value, as described by the localized $4f$ model.
Furthermore, high resolution thermal expansion measurements for the diluted systems in small $B$ have revealed the divergent behavior of a quadrupolar Gr\"{u}neisen ratio, i.e, $\Gamma_{\mathrm{u}} \sim B^2/[T^2 \mathrm{log}(1/T)]$ \cite{Woerl_2022}.
Since the NFL temperature dependences of $C_{\mathrm{m}}/T$, $\Delta\rho$, $(C_{11}-C_{12})/2$, $\Gamma_{\mathrm{u}}$, and the magnetic field response of $(C_{11}-C_{12})/2$ are consistent with the single-site two-channel quadrupolar Kondo model, it was proposed that the single-site quadrupolar Kondo effect manifests itself in the systems \cite{Yamane_2018b, Yanagisawa_2019, Woerl_2022}.
On the other hand, there is another theoretical calculation with a two-channel Anderson impurities model, implying that these NFL behaviors in the diluted Pr systems are essentially the same as those of the quadrupolar Kondo lattice suggested in the $x = 1$ system \cite{Tsuruta_2022, Tsuruta_2024}.

Since the electric quadrupoles are nonmagnetic, it was considered that the quadrupolar Kondo effect cannot be detected by magnetic measurements.
Since the quadrupolar Kondo effect doesn't need localized magnetic dipolar degrees of freedom, it was considered to difficult to obtain its information by magnetic susceptibility and magnetization measurements.
Indeed, inconsistent results regarding the magnetic susceptibility $\chi_{\mathrm{M}}$ by the quadrupolar Kondo effect have been reported so far.
Cox and Makivic \cite{Cox_Makivic_1994} used non-crossing approximation to calculate $\chi_{\mathrm{M}}$, which follows as $1-B\sqrt{T}$ (B is a constant) with a small prefactor \cite{Cox_1998}.
By contrast, Kusunose $et$ $al.$ \cite{Kusunose_1996} took account of the strong repulsion of conduction electrons at the impurity site to calculate $\chi_{\mathrm{M}}$.
The calculation leads to temperature-dependent $\chi_{\mathrm{M}}$ with a remarkable prefactor.
Experimentally, various NFL temperature dependences have been reported such as $\chi_{\mathrm{M}}(T) \propto 1 - B\sqrt{T}$ for Th(U)Be$_{13}$ \cite{Aliev_1995} and Y(U)Pd$_{3}$ \cite{Seaman_1992}, and $\chi_{\mathrm{M}}(T) \propto -$ln$T$ for Th(U)Ru$_2$Si$_2$ \cite{Amitsuka_1994} and La(Pr)Pb$_3$ \cite{Kawae_2013}.

In this work, we first derived the information regarding the quadrupolar Kondo effect on $\chi_{\mathrm{M}}(T)$ from the precise measurements of magnetization for the quadrupolar Kondo candidate Y$_{1-x}$Pr$_x$Ir$_2$Zn$_{20}$.
We chose a single-crystalline sample with $x = 0.028$, where the Pr ion is sufficiently diluted.
The measurements of magnetization were performed down to 0.1 K in magnetic fields up to 9 T.
As a key finding, we observed that $\chi_{\mathrm{M}}(T)$ at $B =0.5$ T is constant for $0.1 < T < 3$ K, retaining the van-Vleck paramagnetism based on the $\Gamma_3$ doublet CEF ground state and $\Gamma_4$ triplet first exited state even in the quadrupolar Kondo regime.
Remarkably, we also found that magnetization divided by magnetic field $M/B$ at $B = 1$ and 2 T exhibits the $-$ln$T$ dependence, which probably reflects $\chi_{\mathrm{Q}}(T)$ through the magnetic-field-induced magnetic moment.


\section{Experimental}

Single-crystalline samples of Y$_{1-x}$Pr$_x$Ir$_2$Zn$_{20}$ were prepared by the Zn-self flux method as described in the previous papers \cite{Yamane_2018a, Yamane_2018b}.
For the initial composition $X$ for Pr, smaller than 0.05, the wavelength dispersive electron-probe microanalysis (EPMA) does not have high resolution.
Therefore, the magnetization value at 1.8 K in $B = 1$ T was used to calculate the real composition $x$ so that the magnetization value matches that calculated using the CEF level scheme for $x = 1$ \cite{Iwasa_2013}.
For the crystals prepared with $X = 0.05$ and 0.03, respectively, the real compositions $x$ were estimated as 0.028  and 0.022.
The magnetization was measured at 1.8--300 K in magnetic fields of $B \leq 5$ T by using a commercial SQUID magnetometer (Quantum Design, MPMS).
The magnetization for $0.1 < T < 1$ K and at 4.2 K was measured in magnetic fields up to 9 T by the Faraday method with a capacitive magnetometer \cite{Sakakibara_1994}, mounted on a $^3$He-$^4$He dilution refrigerator.
A magnetic-field gradient of 10 T/m was applied to the sample.
For all measurements, the magnetic field was applied along the crystallographic [100] direction of the single crystalline sample.

Figure 1(a) shows the isothermal magnetization $M(B)$ of Y$_{0.972}$Pr$_{0.028}$Ir$_2$Zn$_{20}$ with 106.1 mg at 4.2 K.
$M_{\mathrm{raw}}$ and $M_{\mathrm{MPMS}}$ data are derived with the capacitive magnetometer and the MPMS, respectively.
The deviation of $M_{\mathrm{raw}}$ from $M_{\mathrm{MPMS}}$ is ascribed to the background magnetization  $M_{\mathrm{BG}}$.
The magnitude of  $M_{\mathrm{BG}}$ is about 10$^{-3}$ emu in the temperature and magnetic field regions, which order is comparable to the intrinsic magnetization signal of the sample.
Thereby, we estimated $M_{\mathrm{BG}}$ by measuring  a ferromagnetic nickel sheet which was set on the sample stage of the magnetometer in advance (see the Supplemental Material for details \cite{Supll}).
After subtracting $M_{\mathrm{BG}}$ from $M_{\mathrm{raw}}$, the resulting isothermal magnetization $M$ of Y$_{1-x}$Pr$_x$Ir$_2$Zn$_{20}$ for $x =0.028$ at 4.2 K are plotted in Fig. 1(a) as blue closed circles.
The $M(B)$ data agree well with the $M_{\mathrm{MPMS}}(B)$ data in the range of $0.4 \leq B \leq 5$ T.
The disagreement for $B<0.3$ T is due to the low resolution of the capacitive magnetometer in the low field region \cite{Sakakibara_1994}.
The temperature dependences of $M$ were obtained in the same procedure as that of $M(B)$.

\begin{figure}
\center{\includegraphics[width=\linewidth]{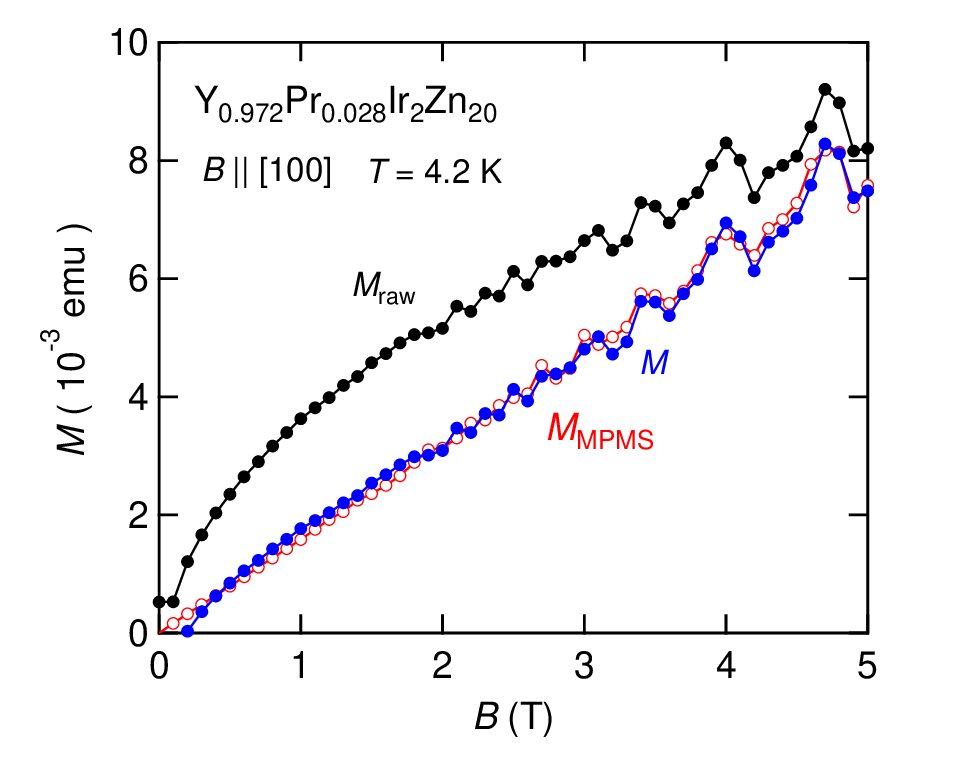}}
\caption{\label{fig:1} 
Magnetization data of Y$_{0.972}$Pr$_{0.028}$Ir$_2$Zn$_{20}$ with 106.1 mg at 4.2 K in magnetic fields of $B \leq 5$ T.
$M_{\mathrm{raw}}(B)$ and $M_{\mathrm{MPMS}}(B)$ were measured with a high-resolution capacitive magnetometer and MPMS, respectively.
$M(B)$ was evaluated by subtracting the background contribution from the $M_{\mathrm{raw}}$ data. Details are described in the main text.
}
\end{figure}

\section{Results and discussion}

First, we discuss the valence and on-site interaction of a Pr ion in Y$_{0.972}$Pr$_{0.028}$Ir$_2$Zn$_{20}$.
Figure 2 shows the temperature dependence of the magnetic susceptibility $\chi_{\mathrm{M}}$ (left-hand-scale) and the inverse $1/\chi_{\mathrm{M}}$ (right-hand-scale) of Y$_{0.972}$Pr$_{0.028}$Ir$_2$Zn$_{20}$ in the magnetic field of $B = 0.5$ T.
The contribution of the Pr ions, $\chi_{\mathrm{M}}^{\mathrm{(Pr)}}$, was estimated as $\chi_{\mathrm{M}}^{\mathrm{(Pr)}} = \chi_{\mathrm{M}} - \chi_{\mathrm{M}}^{\mathrm{(Y)}}$, where $\chi_{\mathrm{M}}^{\mathrm{(Y)}}$ is the magnetic susceptibility of the counterpart YIr$_2$Zn$_{20}$ plotted with the open squares.
The $1/\chi_{\mathrm{M}}^{\mathrm{(Pr)}}(T)$ data for 50--300 K follow the Curie-Weiss law.
The effective magnetic moment $\mu_{\mathrm{eff}}$ of 3.74 $\mu_{\mathrm{B}}$/Pr is close to 3.58 $\mu_{\mathrm{B}}$ for a free Pr$^{3+}$ ion.
The paramagnetic Curie temperature was evaluated to be $\theta_{\mathrm{p}} = -7.04(1)$ K.
The value of $\theta_{\mathrm{p}}$ tends to decrease with decreasing $x$ from $-2.6(2)$ K for $x = 1$ to $-3.0(2)$ K for $x = 0.44$ (not shown), and to $-7.04(1)$ K for $x = 0.028 $.
The larger value of $|\theta_{\mathrm{p}}|$ for $x = 0.028$ suggests enhancement of on-site antiferromagnetic interaction by possible hybridization between the $4f^2$ and conduction electrons.
The enhancement of $|\theta_{\mathrm{p}}|$ by Pr$^{3+}$-dilution possibly arises from the on-site (Kondo) effect pronounced by suppression of inter-site magnetic interaction and/or the positive chemical pressure effect by substituting Pr$^{3+}$ by Y$^{3+}$.
On cooling below 20 K, the $\chi_{\mathrm{M}}^{\mathrm{(Pr)}}(T)$ data approach a constant value, indicating a van-Vleck paramagnetic state.
The van-Vleck paramagnetic behavior is expected for the non-Kramers $\Gamma_3$ doublet CEF ground state as proposed by the previous specific heat and ultrasonic measurements for the diluted samples \cite{Yamane_2018a, Yanagisawa_2019}.

\begin{figure}
\center{\includegraphics[width=\linewidth]{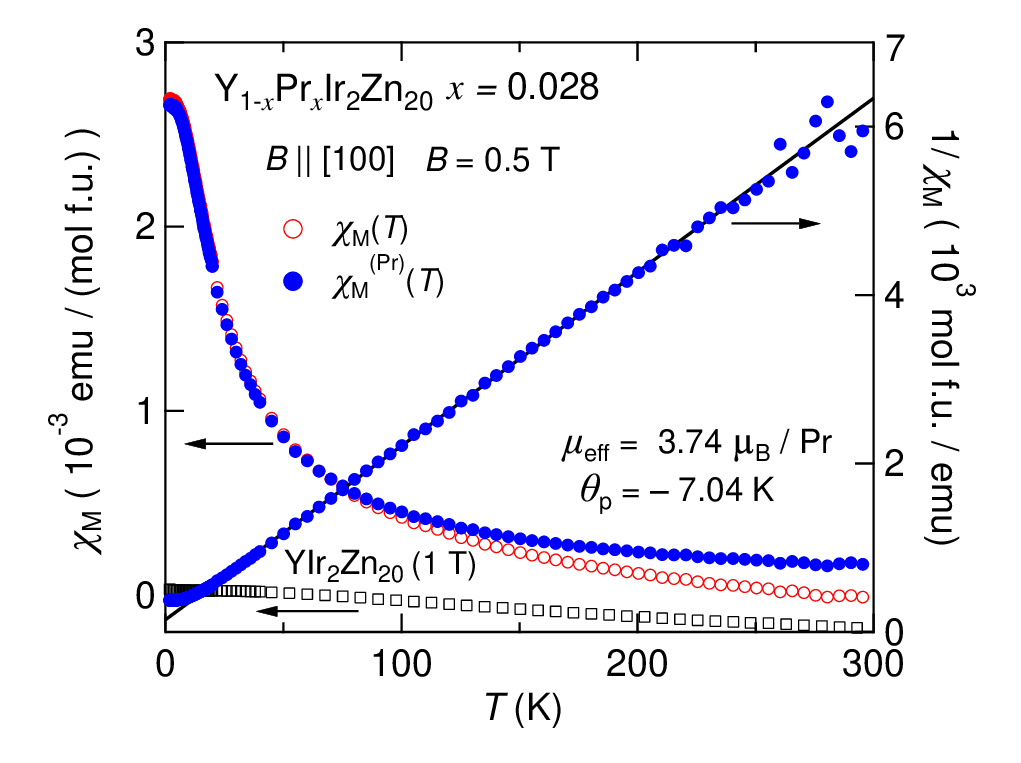}}
\caption{\label{fig:2}Temperature dependence of the magnetic susceptibility $\chi_{\mathrm{M}}$ and the reciprocal $1/\chi_{\mathrm{M}}$ of Y$_{0.972}$Pr$_{0.028}$Ir$_2$Zn$_{20}$ in a magnetic field of $B = 0.5$ T.
Open and closed circles, respectively, describe the raw data and those in which the contribution from the nonmagnetic host metal YIr$_2$Zn$_{20}$ was subtracted.
Open squares are the $\chi_{\mathrm{M}}$ data for YIr$_2$Zn$_{20}$ measured at $B = 1$ T.}
\end{figure}

Next, we show that the low-temperature isothermal magnetization is influenced by CEF effect and quantum oscillation, but those in $1<B<4$ T cannot be explained by only both contributions.
Figure 3 shows the isothermal magnetization $M(B)$ at $T = 0.2$, 0.3, 0.9, and 4.2 K.
All the $M(B)$ curves oscillate as a function of the magnetic fields above 2 T, which oscillations arise from the de Haas-van Alphen (dHvA) effect.
The observation of the dHvA oscillation ensures high quality of the single-crystalline sample.
As shown in the inset of Fig. 3, the dHvA spectrum is derived from a fast Fourier transformation (FFT).
Before performing the FFT, the oscillating components were extracted by subtracting the results of fourth-order polynomial fitting from the raw data in advance.
Due to the small maximum magnetic field and rough intervals of the measurements, we focus on only three dHvA frequencies indicated by arrows.
Among the signals at 26, 53, and 76 T, the latter two are the second and third harmonics of the first one.
According to the first-principle calculation of the band structure for the isostructural compound LuIr$_2$Zn$_{20}$ \cite{Matsushita_2011}, the dHvA oscillation with 26 T would originate from the smallest Fermi surface labeled the branch $\gamma$ with a small effective electron mass of $m^* \sim 0.21m_0$, where $m_0$ is mass of a free electron.

The thick solid curves of $M(B)$ in the main frame of Fig. 3 are calculated by the Zeeman term and the cubic CEF Hamiltonian as described in the following equations \cite{LLW},
\begin{equation}
\mathcal{H}= g_J\mu_{\mathrm{B}}\bm{J}\cdot \bm{B} + \mathcal{H}_{\mathrm{CEF}},
\end{equation} 
\begin{equation}
\mathcal{H}_{\mathrm{CEF}}=W\Bigl[\frac{Y}{60}(O_4^0 + 5 O_4^4)+\frac{1-\left| Y\right| }{1260}(O_6^0 - 21O_6^4)\Bigr].
 \end{equation} 
$g_J$ is the Land$\grave{\mathrm{e}}$ $g$ factor, $\mu_{\mathrm{B}}$ is the Bohr magneton, $W$ and $Y$ are the CEF parameters, and $O_n^m$'s  are the Stevens operators \cite{Stevens_1952} for the total angular momentum of $J = 4$ for a Pr$^{3+}$ ion. 
With the CEF parameters of $W = -1.22$ K and $Y = 0.537$ of PrIr$_2$Zn$_{20}$ \cite{Iwasa_2013} described in the introduction, the CEF ground state is the $\Gamma_3$ doublet separated from the first exited $\Gamma_4$ triplet by 28 K.
As shown with the black solid line, the $M(B)$ data at $T = 4.2$ K except for the dHvA oscillation are reproduced well by the calculation, whereas the $M(B)$ data at 0.2 K in the range of $1< B \leq 4$ T are smaller than the calculated one.
This fact will be discussed in connection with the unusual temperature dependence of $M(T)/B$ in this field range.

\begin{figure}
\center{\includegraphics[width=\linewidth]{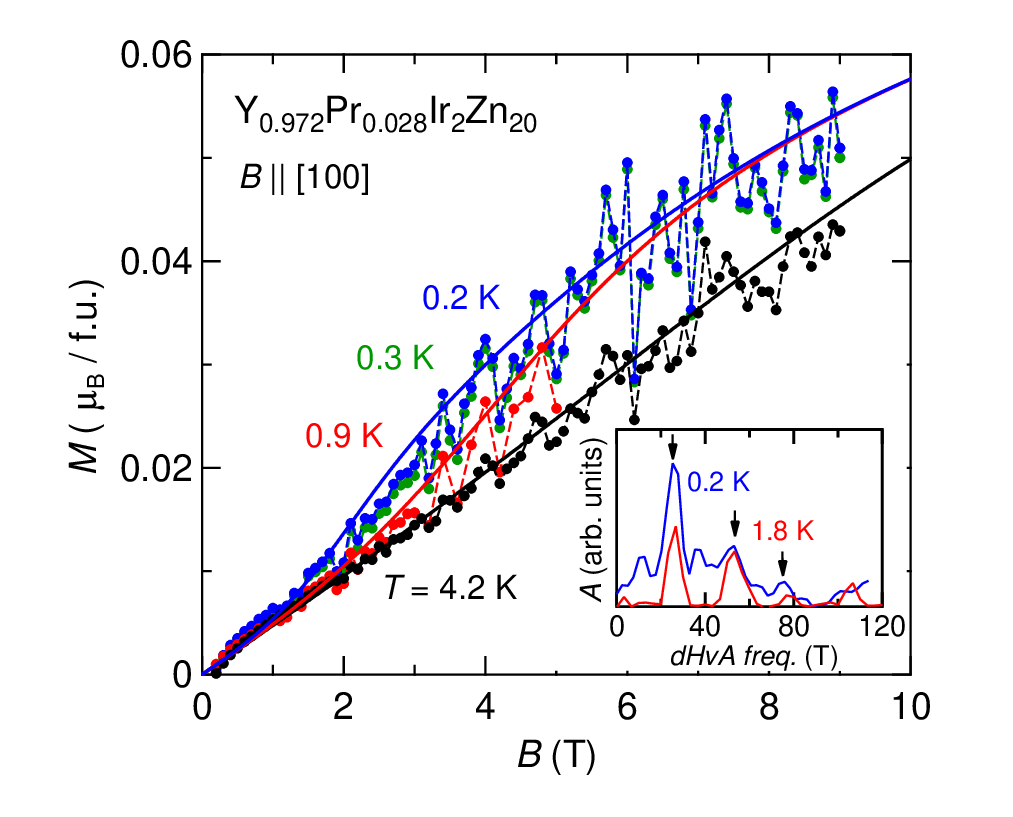}}
\caption{\label{fig:3}Isothermal magnetization curves $M(B)$ of Y$_{0.972}$Pr$_{0.028}$Ir$_2$Zn$_{20}$ at $T = 0.2$, 0.3, 0.9, and 4.2 K in the magnetic fields.
The dashed lines connecting the experimental data are guides to the eyes.
The thick solid curves are calculations with the Hamiltonian of the CEF and Zeeman terms.
See text for detail.
The inset shows fast Fourier transformation spectra for the de Haas-van Alphen oscillations in $M(B)$.}
\end{figure}

The most remarkable result in this work is the magnetic-field variation of the temperature dependence of magnetization.
Figure 4 shows the temperature dependences of the magnetization divided by the magnetic field, $M(T)/B$, of Y$_{0.972}$Pr$_{0.028}$Ir$_2$Zn$_{20}$ at $B = 0.5$, 1, 2, and 4 T.
The data below 1 K contain contributions not only from the Pr$^{3+}$ ions but also from the dHvA oscillations.
To subtract the latter contribution, the data for each magnetic field below 1 K are vertically shifted so that the values at $T$ = 0.2 K agree with those derived by fitting the $M(B)$ curve at 0.2 K with a polynomial function (see the Supplemental Material for details \cite{Supll}).
At $B = 0.5$ T, the $M(T)/B$ data are almost independent of temperature for $0.1 \leq T \leq 3$ K.
On the other hand, at $B = 1$ and 2 T, the $M(T)/B$ data exhibit the $-$ln$T$ dependence from 1 to 0.1 K.
At a higher field of $B = 4$ T, the $M(T)/B$ data approach a constant on cooling below 0.3 K.

We calculate $M(T)/B$ for the four different fields with Eq. (1) and plotted the results by dashed lines.
At $B = 0.5$ T, the constant behavior for $0.1 \leq T \leq 3$ K is reproduced by the calculated line which is based on the van-Vleck paramagnetic behavior expected from the $\Gamma_3$ doublet CEF ground state.
When we regard the data of $M/B$ at $B = 0.5$ T as $\chi_{\mathrm{M}}(T)$, the van-Vleck paramagnetic behavior agrees with the small prefactor of the temperature variation predicted by Cox and Makivic \cite{Cox_Makivic_1994}.
This paramagnetic behavior is much distinct from the NFL temperature dependences for the other U- or Pr-based compounds described in the introduction.
For the U-systems, due to the uncertainty of the valence of U ion, there should be a contribution from the magnetic moments of the Kramers U$^{5+}$ ion ($5f^1$ configuration).
This contribution can enhance the magnetic susceptibility.
For La(Pr)Pb$_3$, the linear term of the magnetic susceptibility against magnetic field has been derived by fitting the $M/B$ versus $B^2$ data for $B > 1$ T to avoid the change by the superconductivity of Pb films \cite{Kawae_2013}.
As discussed in Ref. \cite{Kawae_2013} and the next paragraph, the $M/B$ data of both La(Pb)Pb$_3$ and Y$_{0.972}$Pr$_{0.028}$Ir$_2$Zn$_{20}$ above 1 T may be modified by the contribution of the first-excited $\Gamma_4$ triplet.

\begin{figure}
\center{\includegraphics[width=\linewidth]{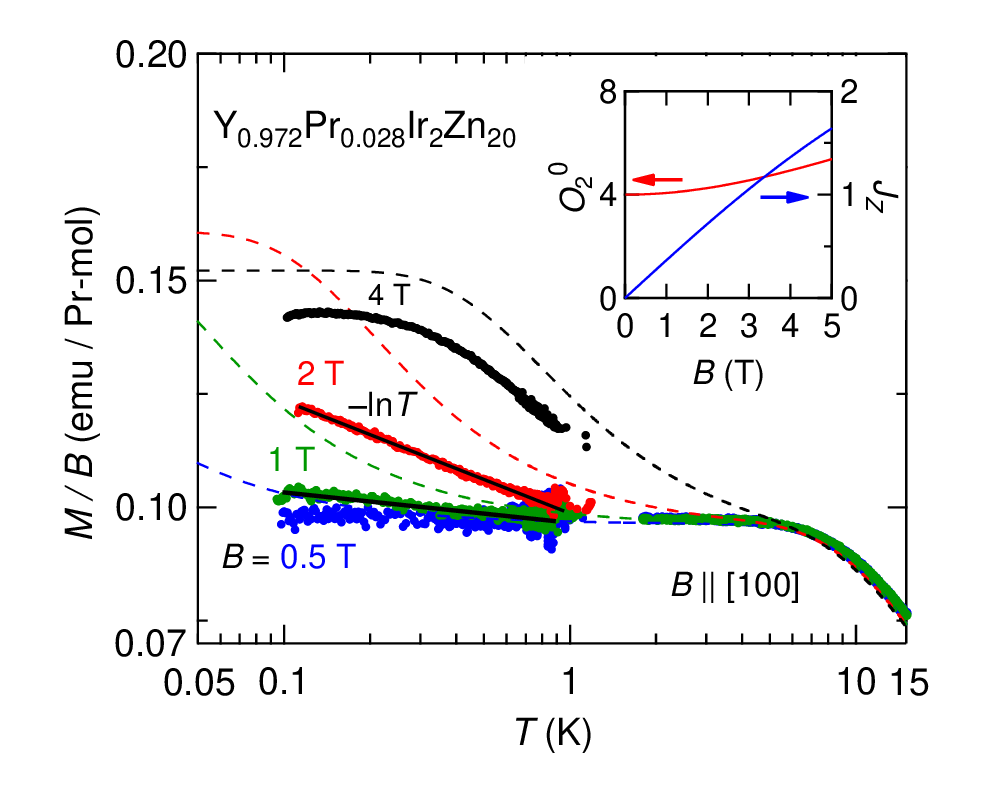}}
\caption{\label{fig:4} Temperature dependence of the magnetization divided by the magnetic field, $M(T)/B$, of Y$_{0.972}$Pr$_{0.028}$Ir$_2$Zn$_{20}$.
The magnetic fields of 0.5, 1, 2, and 4 T were applied along the [100] direction.
Dashed curves are calculated with the CEF and the Zeeman terms.
See text for details.
The inset shows the magnetic field ($B$ $||$ [100]) variations of the mean values of the magnetic dipolar operator $J_Z$ and the electric quadrupolar operator $O_2^0$ of the ground state, respectively.}
\end{figure}

By contrast, at $B = 1$ and 2 T, the calculated lines become larger than the experimental ones as temperature is decreased below 1 K, then do not reproduce the observed $-$ln$T$ dependence.
We here discuss the three scenarios for the $-$ln$T$ dependence emerging at $B = 1$ and 2 T.
First, the ordinary Kondo effect by magnetic impurities in the sample could lead a $-$ln$T$ dependence.
However, no magnetic impurity was detected by the EPMA.
In addition, the impurity contribution should become remarkable as $B$ decreases.
This is not the case because the $M(T)/B$ data at $B = 0.5$ T are independent of temperature between 0.1 and 3 K.
Therefore, the contribution of the magnetic impurity can be ruled out.
Second, the nuclear magnetic moment of $^{141}$Pr could contribute to the total magnetization.
The nuclear magnetic moment should increase with increasing $B$ or decreasing $T$. 
This contradicts with the observation that the $M(T)/B$ data at 4 T approach a constant for $T < 0.3$ K.
Third, we consider the magnetic field variations of the magnetic dipole $J_Z$ and the electric quadrupole $O_2^0$ of the ground state of Pr$^{3+}$ ion in Y$_{0.972}$Pr$_{0.028}$Ir$_2$Zn$_{20}$.
The contribution of $J_Z$ and $O_2^0$ to the magnetization in magnetic fields can be derived by calculating the mean values of the magnetic dipolar operator $\hat{J_Z}$ and the electric quadrupolar operator $\hat{O_2^0}$ with the following equations
\begin{equation}
\begin{split}
J_Z (B) &= \bra{\psi_{\mathrm{GS}}(B)}\hat{J_Z}\ket{\psi_{\mathrm{GS}}(B)}, \\
O_2^0 (B) &= \bra{\psi_{\mathrm{GS}}(B)}\hat{O_2^0}\ket{\psi_{\mathrm{GS}}(B)},
\end{split}
\end{equation} 
where $\psi_{\mathrm{GS}}(B)$ is the wave function of the lowest lying CEF state in magnetic fields calculated with Eq. (1).
The inset of Fig. 4 shows the mean values of $\hat{J_Z}$ and $\hat{O_2^0}$ calculated for the lowest-lying CEF state as a function of the magnetic field.
The $\Gamma_3$ doublet is split quadratically by magnetic fields through the off-diagonal components of the Zeeman term mainly between the CEF ground $\Gamma_3$ doublet and the first excited $\Gamma_4$ triplet.
The splitting of the $\Gamma_3$ doublet results in field-induced values of $J_Z$, which gives rise to temperature dependent $\chi_{\mathrm{M}}$.
Thereby, we speculate that the $-$ln$T$ dependence of $M(T)/B$ at $B = 1$ and 2 T reflects the susceptibility of overscreened quadrupoles in the quadrupolar Kondo regime, via field-induced magnetic dipolar moment $J_Z$.
In fact, the ln$T$ dependent softening was observed by the measurements of the $\Gamma_3$-type elastic constant $(C_{11}-C_{12})/2$ at $0 \leq B \leq 2$ T.
This scenario reminds us of quadrupolar ordered structure determination by utilizing neutron diffraction from magnetic-field-induced magnetic moments \cite{Onimaru_2005}.
Therefore, we propose that the overscreened quadrupoles and their fluctuations can be observed by magnetic probes such as nuclear magnetic resonance, ac magnetic susceptibility, and neutron  scattering, through the magnetic-field-induced magnetic moments.

For a higher field of $B = 4$ T, the $M(T)/B$ data approach the constant on cooling below 0.3 K, which behavior is qualitatively reproduced by the calculated data shown with the (black) dashed line.
When the Zeeman splitting energy of the $\Gamma_3$ CEF ground doublet exceeds the energy scale of the quadrupolar Kondo effect, which must be relevant to $T_{\mathrm{K}}$, the field-induced local $4f$ singlet ground state would exhibit van-Vleck paramagnetic behavior.
The reproduction, therefore, indicates the localized $4f$ electronic state induced by magnetic field of $B = 4$ T.
This consideration is supported by the suppression of the ln$T$ softening of $(C_{11}-C_{12})/2$ in $B > 2$ T. \cite{Yanagisawa_2019}.
On the other hand, the experimental $M(T)/B$ values are smaller than the calculated one based on the CEF level scheme of PrIr$_2$Zn$_{20}$.
Because the ionic radius of Y$^{3+}$ ion is smaller than that of Pr$^{3+}$ ion, the CEF splitting energy of the Pr$^{3+}$ ion diluted in YIr$_2$Zn$_{20}$ should be larger than that in PrIr$_2$Zn$_{20}$.
Thereby, enhanced CEF splitting can suppress the van-Vleck term resulting in the suppression of the magnetization value.
In fact, the linear enlargement of lattice parameter with increasing Pr$^{3+}$ ion is reported in Ref. \cite{Yamane_2018a}.
Otherwise, the moderately enhanced on-site antiferromagnetic interaction suggested by $\theta_{\mathrm{p}} = -7.04(1)$ K can suppress the magnetization.

\section{Conclusion}
We have measured the magnetization $M$ of the diluted Pr system Y$_{0.972}$Pr$_{0.028}$Ir$_2$Zn$_{20}$ to reveal the quadrupolar Kondo effect on the temperature dependence of magnetization.
The van-Vleck paramagnetic behavior of $\chi_{\mathrm{M}}$ in 0.1--3 K matches theoretical prediction proposed by Cox and Makivic \cite{Cox_Makivic_1994, Cox_1998}.
The logarithmic temperature dependence of $M(T)/B$ at $B = 1$ and 2 T probably reflects the behavior of the overscreened quadrupoles, through the magnetic-field-induced magnetic moments.
This result suggests that the quadrupolar dynamics can be investigated by measurements such as nuclear magnetic resonance, ac magnetic susceptibility, and quasi-elastic neutron scattering via the magnetic-field-induced magnetic moments.
In a higher field of 4 T, $M(T)/B$ approaches a constant on cooling below 0.3 K.
This remarkable change in $M(T)/B$ occurs when the Zeeman splitting could exceed the energy scale of the quadrupolar Kondo effect to realize the magnetic-field-induced localized state.

\section*{Acknowledgments}

We would like to thank A. Tsuruta, T. Yanagisawa, S. Hoshino, J. Otsuki, H. Kusunose, H. Amitsuka, R. Yamamoto, K. Izawa, H. Harima, K. Miyake, H. Tou, K. Iwasa, T. D. Matsuda for helpful discussion.
We also thank Y. Shibata for the electron-probe microanalysis performed at N-BARD, Hiroshima University.
The measurement of the magnetization with the MPMS and the $^3$He-$^4$He dilution refrigerator were carried out at N-BARD, Hiroshima University.
We acknowledge support from Center for Emergent Condensed-Matter Physics (ECMP), Hiroshima University.
This work was supported by grants from MEXT/JSPS of Japan, Grants Nos. JP26707017, JP15KK0169, JP15H05886, JP16H01076 (J-Physics), JP18H01182, JP23H04870, and JP24K00574.




\newpage

\begin{center}
\vspace{5mm}
{\bf \large Supplemental material for ``Single-site quadrupolar Kondo effect in a diluted non-Kramers doublet system Y$_{1-x}$Pr$_x$Ir$_2$Zn$_{20}$ for $x = 0.028$ viewed from magnetization''}\\
\vspace{5mm}
Yu Yamane, Takahiro Onimaru, Yasuyuki Shimura, Suguru Tsuda, Kazunori Umeo, and Toshiro Takabatake\\
\vspace{5mm}
\end{center}

\renewcommand{\thefigure}{S\arabic{figure}}
\renewcommand{\thesection}{S\Roman{section}}
\renewcommand{\thesubsection}{S\Alph{subsection}}
\setcounter{table}{0}
\setcounter{figure}{0}
\setcounter{section}{0}

In Section SI, we describe how to evaluate the background of the capacitive magnetometer, $M_{\mathrm{BG}}(B, T)$, using the magnetic field and temperature dependences of magnetization for a ferromagnetic nickel sheet. In Section SII, we explain how to subtract the contribution of de Haas-van Alphen effects from the temperature dependences of magnetization.

\section{Background of the capacitive magnetometer}

\subsection{Comparison between the magnetization data derived with the capacitive magnetometer and the MPMS}
We estimated the background contribution $M_{\mathrm{BG}}(B, T)$ at 4.2 K by comparing the magnetization data of Y$_{1-x}$Pr$_x$Ir$_2$Zn$_{20}$ for $x =0.028$ with 106.1 mg derived with the capacitive magnetometer and the MPMS.
Figure S1 shows the isothermal magnetization $M(B)$ at 4.2 K derived with the capacitive magnetometer $M_\mathrm{raw}$ (black closed circle),  and the MPMS, $M_\mathrm{MPMS}$ (red open circle).
The $M_{\mathrm{BG}}$ values at 4.2 K estimated by subtracting the $M_\mathrm{MPMS}$ data from $M_\mathrm{raw}$ are displayed as well with green closed circles.
Assuming a practical fitting function, $M(B) = A_0 ~ + ~ A_1 e^{-B/A_2} + A_3 B$, where $A_i$ $(i = 1, 2, $and $3)$ are coefficients, the $M_{\mathrm{BG}}$ can be fitted as shown with the green solid curve.

\begin{figure}
\center{\includegraphics[width=8cm]{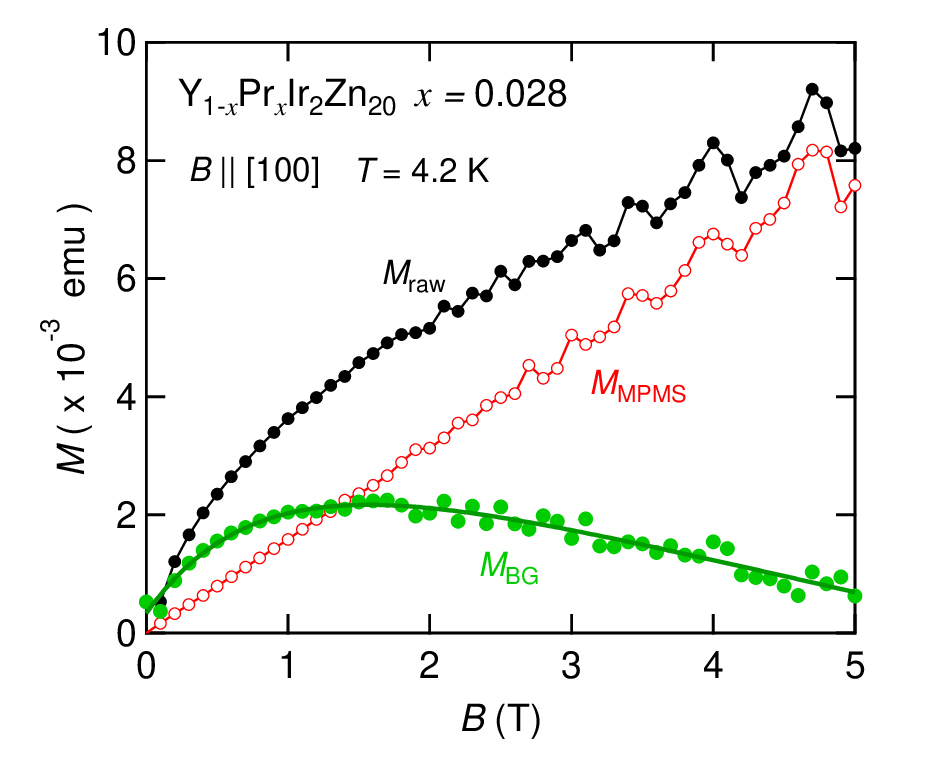}}
\caption{\label{fig:1} 
Isothermal magnetization curves of Y$_{1-x}$Pr$_x$Ir$_2$Zn$_{20}$ for $x =0.028$ at 4.2 K measured with the capacitive magnetometer, $M_{\mathrm{raw}}$ (black closed circles), and the MPMS, $M_{\mathrm{MPMS}}$ (red open circles).
The green closed circles and solid line represent the background data of the capacitive magnetometer and its fitting result, respectively (see main text for detail).
}
\end{figure}

\subsection{Background evaluation from the magnetization measurements of a ferromagnetic nickel sheet}
To evaluate $M_{\mathrm{BG}}$ for $T < 4.2$ K, we measured the magnetization of a ferromagnetic nickel sheet with 0.56 mg by the capacitive magnetometer.
Based on the data presented in Fig. S2, the $M(T)$ data for the nickel sheet is independent of temperature below 10 K.
It is because the ferromagnetic order parameter with the Curie temperature $T_{\mathrm{C}} = 631$ K \cite{Legendre_2011} is already well-developed at such low temperatures.
Thus, to evaluate $M_{\mathrm{BG}}$, the $M(B)$ data measured with the MPMS must be subtracted from those obtained using the capacitive magnetometer.
Bearing this in mind, we measured the magnetic field and temperature dependences of magnetization of the nickel sheet with 0.56 mg by utilizing both the capacitive magnetometer and the MPMS.

Figure S3 shows the $M(B)$ data for the nickel sheet, where the closed and open circles are the data measured with the capacitive magnetometer and the MPMS, respectively.
We obtained the differences between the data except for $B < 2$ T as the background contribution.
It is noted that the difference for $B < 2$ T possibly originates from the small magnetic torque due to the anisotropic shape effect of the sheet-like sample.
Thus, assuming the function derived in Sec. {\bf SA}, we fitted the $M(B)$ curves.
The result is displayed in Fig. S4.
After showing a maximum at about 2 T, the values of $M_{\mathrm{BG}}$ decrease monotonically with increasing $B$.
The maximum probably results from magnetic impurities of the copper stage in the capacitive magnetometer and the decrease is ascribed to the diamagnetic contribution.
Utilizing the $M_{\mathrm{BG}}(B)$ data, we estimated the intrinsic $M(B)$ of Y$_{1-x}$Pr$_x$Ir$_2$Zn$_{20}$ for $x =0.028$ in the main text.
The temperature dependences of $M$ were estimated as well in the same procedure as the estimation of $M_{\mathrm{BG}}(B)$.

\begin{figure}
\center{\includegraphics[width=8cm]{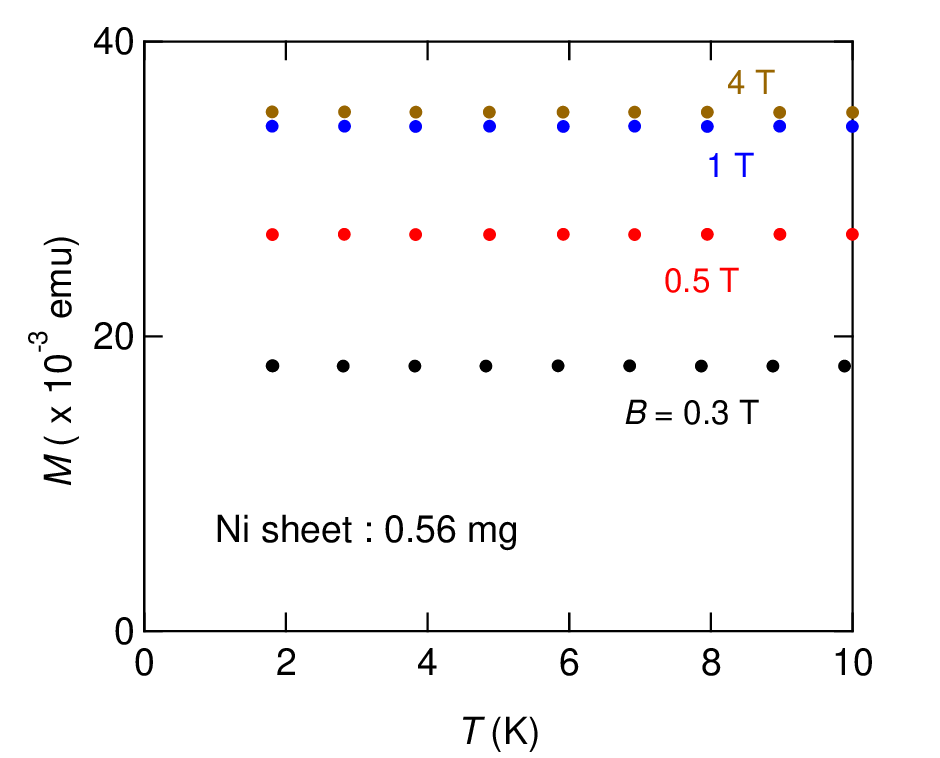}}
\caption{\label{fig:1}
Temperature dependences of magnetization for a nickel sheet below 10 K, which was measured at 0.3, 0.5, 1, and 4 T with the MPMS.
}
\end{figure}

\begin{figure}
\center{\includegraphics[width=8cm]{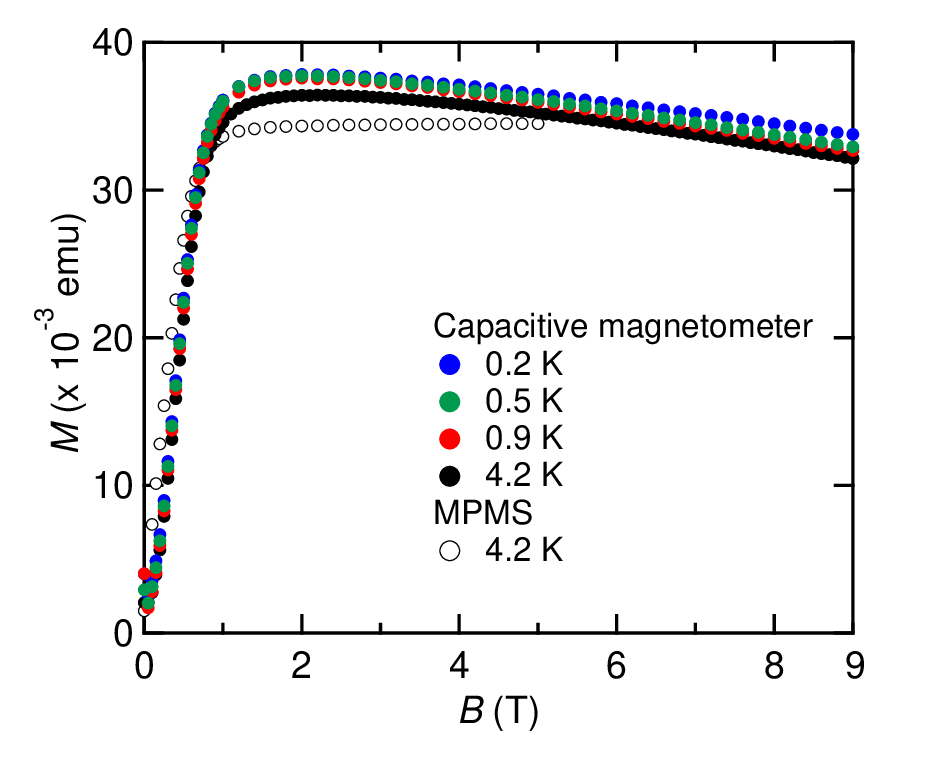}}
\caption{\label{fig:1} 
Isothermal magnetization of the nickel sheet at various temperatures.
The closed and open circles are the data measured with the capacitive magnetometer and the MPMS, respectively.
}
\end{figure}

\begin{figure}
\center{\includegraphics[width=8cm]{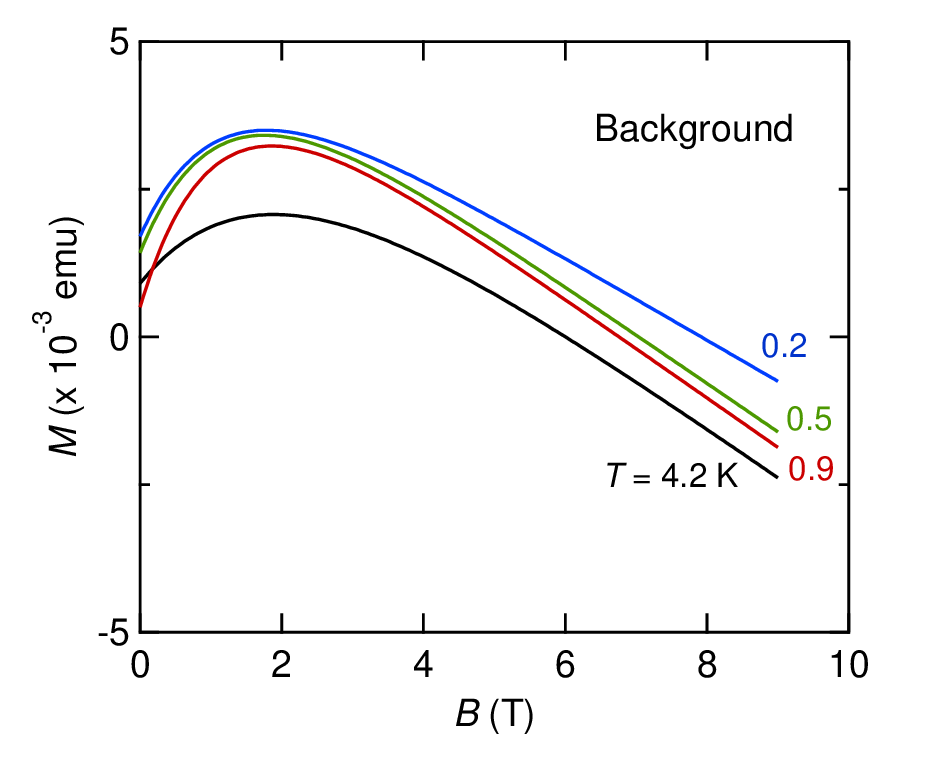}}
\caption{\label{fig:1} 
Magnetic field dependences of the background of the capacitive magnetometer at $T$ = 0.2, 0.5, 0.9, and 4.2 K in magnetic fields of $B \leq 9$ T.
}
\end{figure}

\section{Subtraction of the contribution of de Haas-van Alphen effects to the temperature dependences of magnetization}

The data of temperature dependences of magnetization divided by magnetic field, $M(T)/B$, below 1 K contain contributions not only from the Pr$^{3+}$ ions but also from dHvA oscillations.
Then, we estimate the contribution from the dHvA effect as follows.
Figure S5 shows the magnetization curve $M(B)$ of Y$_{1-x}$Pr$_x$Ir$_2$Zn$_{20}$ for $x =0.028$ at 0.2 K, where the magnetic fields are applied along the [100] axis.
The $M(B)$ curve above 1 T oscillates with increasing $B$ due to the dHvA effect.
We fitted the $M(B)$ data in the range of $0.4 < B < 5$ T with an odd-numbered polynomial function $M(B)$ = $C_1B$ + $C_3B^3$, where $C_1$ and $C_3$ are coefficients, and plotted in Fig. S5 with the blue solid curve.
Assuming that the difference between the fitting curve and the $M(B)$ data is attributed to the dHvA effect, we vertically shifted the $M(T)/B$ data below 1 K shown in Fig. 4 so that the values at $T$ = 0.2 K agree with those derived by the fitting curve.

\begin{figure}
\center{\includegraphics[width=8cm]{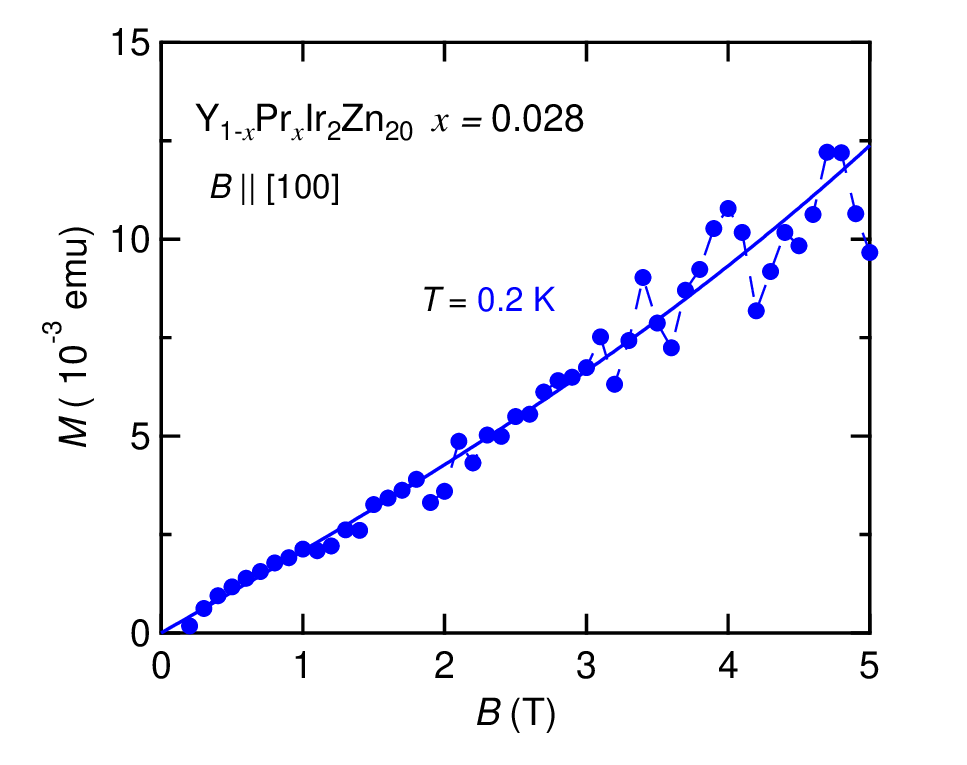}}
\caption{\label{fig:1} 
Fitting result for magnetization curve of Y$_{1-x}$Pr$_x$Ir$_2$Zn$_{20}$ for $x =0.028$ with polynomial function.
}
\end{figure}




\end{document}